\documentclass[aps,prb,reprint,twocolumn,superscriptaddress]{revtex4-2}

\usepackage{graphicx}
\usepackage{bm}
\usepackage{amsmath}  
\usepackage{amssymb}
\usepackage{hyperref}
\usepackage{xcolor}
\usepackage{array}
\usepackage{multirow}
\usepackage{physics}
\usepackage{lipsum}
\usepackage[T1]{fontenc}

\renewcommand{\v}[1]{{\boldsymbol{#1}}}%
\newcommand{\ov}{\overline}

\begin{document}
\title{Magic-angle helical trilayer graphene} 
\author{Trithep Devakul}
\email{tdevakul@mit.edu}  
\affiliation{Department of Physics, Massachusetts Institute of Technology, Cambridge, MA 02139, USA}
\author{Patrick J. Ledwith}
\affiliation{Department of Physics, Harvard University, Cambridge, MA 02138, USA}
\author{Li-Qiao Xia}
\affiliation{Department of Physics, Massachusetts Institute of Technology, Cambridge, MA 02139, USA}
\author{Aviram Uri}
\affiliation{Department of Physics, Massachusetts Institute of Technology, Cambridge, MA 02139, USA}
\author{Sergio de la Barrera}
\affiliation{Department of Physics, Massachusetts Institute of Technology, Cambridge, MA 02139, USA}
\affiliation{Department of Physics, University of Toronto, Toronto, ON M5S 1A7, Canada}
\author{Pablo Jarillo-Herrero}
\affiliation{Department of Physics, Massachusetts Institute of Technology, Cambridge, MA 02139, USA}
\author{Liang Fu}
\affiliation{Department of Physics, Massachusetts Institute of Technology, Cambridge, MA 02139, USA}
\begin{abstract}

We propose helical trilayer graphene (HTG), a helical structure featuring identical rotation angles
$\theta\approx 1.5^\circ$
between three consecutive layers of graphene, as a unique and experimentally accessible platform for realizing exotic correlated topological states of matter.
While nominally forming a supermoir\'e (or moir\'e-of-moir\'e) structure, we show that HTG locally relaxes into large regions of a periodic single-moir\'e structure in which $C_{2z}$ is broken, giving rise to flat topological bands carrying valley-Chern numbers $C=\pm(1,-2)$.
These bands feature near-ideal quantum geometry
and are isolated from remote bands by a large gap $E_{\mathrm{gap}}\sim 100$ meV, making HTG a promising platform for experimental realization of correlated topological states such as integer and fractional quantum anomalous Hall states in $C=1$ and $2$ bands.
\end{abstract}
\maketitle

The intricate interplay of topology and strong electronic interactions is one of the most fascinating and rapidly evolving areas of modern condensed matter physics.
Following the discovery of superconductivity and strong correlations in twisted bilayer graphene (TBG) 
\cite{cao2018unconventional,cao2018correlated},
moir\'e materials 
 have risen to the forefront of both theoretical and experimental condensed matter physics research as an ideal platform for exploring strongly correlated physics in topological bands~\cite{andrei2021marvels}.
In the graphene family, significant progress has also been made in multilayer moir\'e heterostructures, such as alternating twist multilayers~\cite{khalafMagicAngleHierarchy2019,park2021tunable,hao2021electric,cao2021pauli,park2022robust,zhang2022promotion}, or single twist multilayers~\cite{polshyn2020electrical,chen2021electrically,xu2021tunable,he2021competing,cao2020tunable,liu2020tunable,he2021symmetry,heChiralitydependentTopologicalStates2021} such as twisted monolayer-bilayer graphene.
In parallel, moir\'e heterostructures based on semiconductor transition metal dichalcogenides (TMD) have also revealed a trove of complementary physics ranging from generalized Wigner crystals to topological states~\cite{mak2022semiconductor}.
The sheer versatility of the moir\'e platform has lead to the experimental realization of an extraordinarily diverse array of physical phenomena.

In magic-angle TBG, 
a manifold of nearly flat isolated single-particle bands enables a unique regime of physics dominated by interactions and band geometry.
Perhaps the most fascinating and direct observations of strongly correlated topology are the quantum anomalous Hall (QAH)\cite{serlinIntrinsicQuantizedAnomalous2020,zhangTwistedBilayerGraphene2019,bultinckMechanismAnomalousHall2020,xieNatureCorrelatedInsulator2020} and fractional Chern insulator (FCI)\cite{xieFractionalChernInsulators2021,ledwithFractionalChernInsulator2020,abouelkomsanParticleHoleDualityEmergent2020,repellinChernBandsTwisted2020,wilhelmInterplayFractionalChern2021,parkerFieldtunedZerofieldFractional2021}, lattice analogues of the integer and fractional quantum Hall states driven by intrinsic band geometry rather than Landau level physics \cite{parameswaranFractionalQuantumHall2013,bergholtzTOPOLOGICALFLATBAND2013,liuRecentDevelopmentsFractional2022,neupertFractionalQuantumHall2011,shengFractionalQuantumHall2011,regnaultFractionalChernInsulator2011,scaffidiAdiabaticContinuationFractional2012,royBandGeometryFractional2014,kourtisFractionalChernInsulators2014}.
However, these topological states in TBG are often fragile and overpowered by competing non-topological states, likely because they require hBN-alignment \cite{sharpeEmergentFerromagnetismThreequarters2019,serlinIntrinsicQuantizedAnomalous2020} or spontaneous breaking of $C_{2z} \mathcal{T}$ symmetry \cite{stepanovCompetingZeroFieldChern2021}.
The FCI states have thus far only been observed in a substrate aligned sample and at finite magnetic field $B\sim 5$T\cite{xieFractionalChernInsulators2021}.
The apparent requirement of substrate alignment poses a significant experimental challenge that severely limits reproducibility of strongly correlated topology in the TBG platform, and it is not clear whether the FCI state can be made stable at zero field.
Very recently, evidence of a zero-field FCI was found in a twisted TMD homobilayer~\cite{cai2023signatures,zeng2023integer}.
It is therefore an important theoretical task to identify new platforms in which such topological states may appear most robustly, as well as for the realization of further exotic phases of matter.

We propose ``helical trilayer graphene'' (HTG), a helical structure featuring identical rotation angles between three consecutive layers of graphene, as a promising and experimentally accessible platform for realizing exotic topological states of matter.
As we will elaborate, unrelaxed HTG does not realize a single periodic moir\'e superlattice, but instead realizes a supermoir\'e (or ``moir\'e-of-moir\'e'')  structure~\cite{zhu2020twisted,mao2023supermoire,zhang2021correlated}).
Nevertheless, we show that HTG locally relaxes into large regions hosting a single-moir\'e structure featuring a periodic honeycomb lattice of the AA stacking regions (shown in Figs~1a,b).
In these regions, which we call h-HTG, $C_{2z}$ is broken by a fixed lateral shift $\v d=\pm\v \delta$ between the two moir\'e superlattices. %
Remarkably, we find that at a magic angle $\theta\approx1.5^\circ$, the moir\'e band structure of h-HTG features a pair of flat, isolated, nearly-degenerate topological bands, shown in Fig2b, with valley-contrasting Chern numbers $C=\pm(1,-2)$.  
Since each valley carries a net Chern number, even the band insulators are topological quantum valley-Hall states, and valley polarization alone yields a net Chern number. 
In particular, these bands feature remarkably uniform charge and Berry curvature distributions, as well as ``near-ideal quantum geometry"\cite{ledwith2022vortexability,ledwithFamilyIdealChern2022,ledwithFractionalChernInsulator2020,dong2022exact,wangExactLandauLevel2021,ledwithStrongCouplingTheory2021,parkerFieldtunedZerofieldFractional2021,gao2022untwisting,varjas2022}, making HTG a promising platform for realizing FCI states in $\abs{C}=1$ and $2$ Chern bands.
Furthermore, the topological flat band manifold is isolated from remote bands by a very large gap $E_{\mathrm{gap}}\sim 100$ meV, implying a high degree of stability and providing a potential route to higher temperature QAH and FCI states.

Zooming out, HTG realizes large regions of h-HTG domains (and its $C_{2z}$ related counterpart), which form a triangular tiling on the supermoir\'e scale, as shown in Fig 1a.
These domains are large (several hundreds of nanometers) so the bulk properties of h-HTG are accessible via local probes such as scanning single-electron transistors~\cite{xieFractionalChernInsulators2021,yu2022correlated,yu2022spin} and scanning nano superconducting quantum interference devices~\cite{grover2022chern}.
Furthermore, the domain size may be tuned via heterostrain engineering~\cite{huder2018electronic,bi2019designing}, and with a small amount of uniform heterostrain ($\approx 0.03\%$), the entire device can relax into a single domain of h-HTG, providing a route to a quantized Hall response measurable by transport.

\begin{figure}[t]
\includegraphics[width=0.5\textwidth]{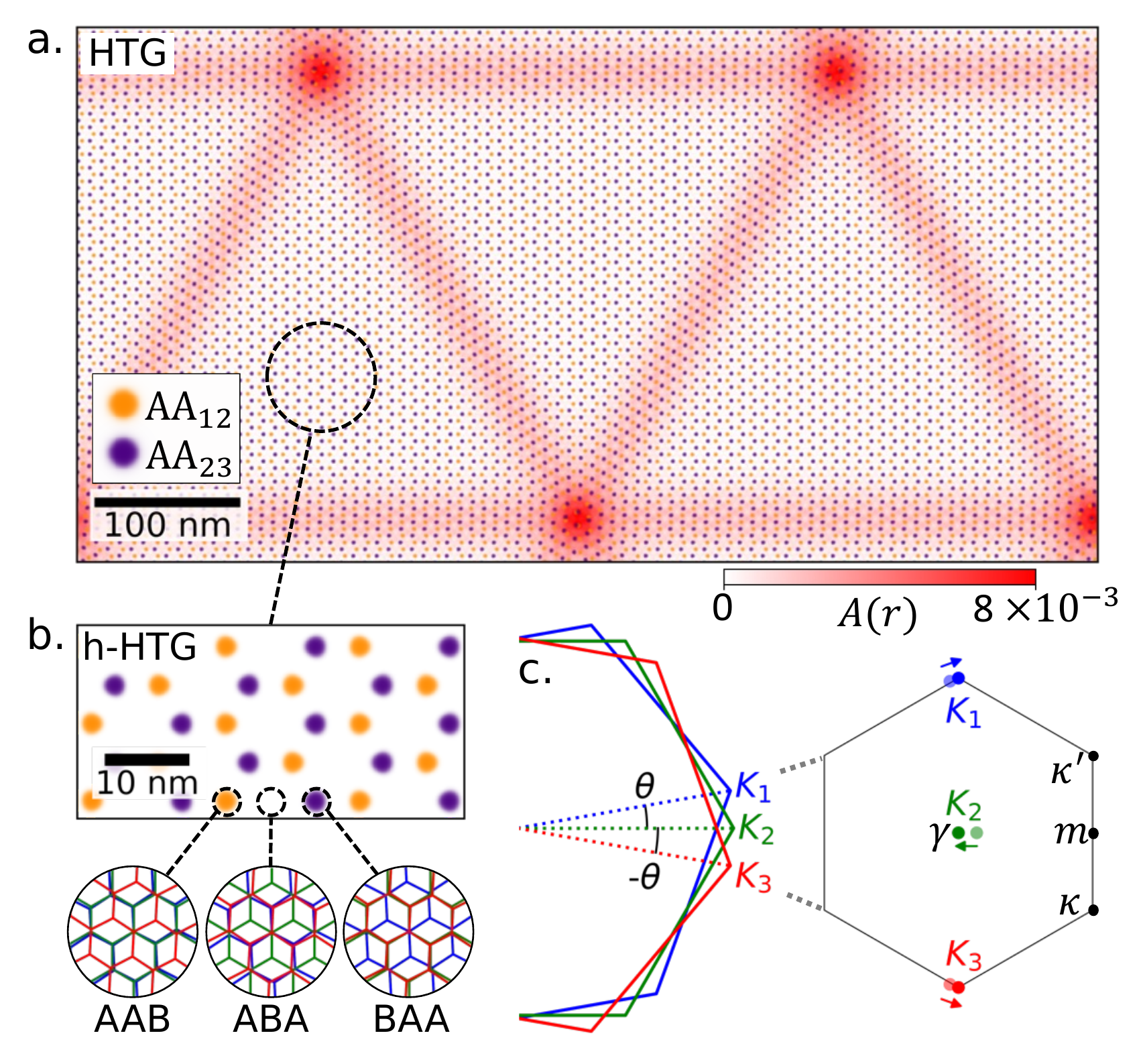}
\caption{
(a) The relaxed structure of HTG at $\theta=1.5^\circ$, where orange and purple dots show the AA stacking regions of adjacent layer pairs, and the red background indicates the moir\'e aperiodicity $A(\v r)$.
The system relaxes to large triangular domains of h-HTG (and its $C_{2z}$ counterpart, $\overline{\mathrm{h}}$-HTG), a periodic moir\'e superlattice with $A(\v r)\approx 0$, separated by a network of domain walls.
(b) A zoom in to the h-HTG region and a further zoomed in illustration of the atomic scale structure at high symmetry points.  
(c) The monolayer graphene BZs for each layer are shown.
In the h-HTG region, the three $K$ points relax onto a single line and fold to the $\kappa,\gamma,\kappa^\prime$ points on the mBZ as illustrated on the right.
}
\end{figure}

When the domain size is finite, a triangular network of interwoven domain walls is realized as shown in Fig1a.
When the domains are tuned to incompressible states at integer or fractional filling, including full and empty filling of the flat bands, the low-energy electronic physics is dominated by the network of gapless domain walls.
This system therefore provides a natural realization of chiral or counter-propagating edge network models~\cite{chalker1988percolation,ho1996models} on the supermoir\'e scale.
Taken together, our work demonstrates that HTG is a uniquely exciting platform for realizing robust strongly correlated topology, gapless edge networks, and for exploring their interplay.

The key ingredient that enables all this richness is lattice relaxation on the supermoir\'e scale,  an aspect which was not fully incorporated in previous theoretical studies.
Refs~\cite{mora2019flatbands,popov2023magic} focused on the electronic properties of a different single-moir\'e superlattice defined by $\v d=0$, which we find is energetically unfavorable and is minimized in the relaxed structure of Fig1a.
Refs~\cite{zhu2020twisted,mao2023supermoire} examined the electronic properties of the full unrelaxed supermoir\'e structure, thus missing the physics of h-HTG.
Various extensions to higher number of layers have also been explored~\cite{wu2020three,liang2022moire,ma2023doubled}.

This paper is structured as follows:  
We first introduce the HTG structure and demonstrate that relaxation favors the formation of a network of large h-HTG domains. 
We examine the electronic properties of h-HTG and its symmetries via an effective continuum model description, revealing the advertised magic angle, topological flat bands, and large remote band gap. 
We then study the model in the ``chiral limit''\cite{tarnopolskyOriginMagicAngles2019}, which features exactly flat bands with ``ideal quantum geometry"\cite{ledwith2022vortexability,ledwithFamilyIdealChern2022,ledwithFractionalChernInsulator2020,dong2022exact,wangExactLandauLevel2021,ledwithStrongCouplingTheory2021,parkerFieldtunedZerofieldFractional2021,gao2022untwisting,varjas2022}, explaining the origin of the magic angle.
Finally, we examine the features that make h-HTG promising for the realization of strongly correlated topology, and discuss possible correlated states at integer and fractional fillings.

\section{Supermoir\'e reconstruction}

We consider the HTG structure consisting of three graphene layers with the twist configuration $(\theta_1,\theta_2,\theta_3)=(\theta,0,-\theta)$. 
In the absence  of lattice relaxation, the moir\'e superlattices of the lower and upper two graphene layers are themselves misaligned by an angle $\theta$, which therefore forms a supermoir\'e structure.
This results in a parametric separation of lengthscales:
the atomic lengthscale $a_0=2.46$\AA{}  is much smaller than the moir\'e lengthscale $a_m= a_0/(2\sin\frac{\theta}{2})$ which is in turn much smaller than the supermoir\'e lengthscale $a_{mm}=a_0/(2\sin\frac{\theta}{2})^2$.

Because of the large supermoir\'e lengthscale, lattice relaxation plays a pivotal role in the physics of HTG and cannot be ignored.
This is because even a small amount of atomic lattice relaxation can result in a magnified effect on the moir\'e scale, and hence a doubly magnified effect at the supermoir\'e scale.
As an analogy, consider the bilayer case.
In TBG at $\theta\approx1.1^\circ$, lattice relaxation is minor and typically accounted for by a phenomenological parameter $\kappa$ known as the chiral ratio.
At very small angles $\theta\lesssim 1^\circ$ ($a_m\gtrsim 14$ nm), however, lattice relaxation results in severe moir\'e lattice reconstruction~\cite{yoo2019atomic,mcgilly2020visualization,alden2013strain,dai2016twisted,nam2017lattice,carr2018relaxation}: the energetically favorable AB and BA stacking regions are enlarged to form large triangular domains of locally atomically-periodic Bernal stacking regions at the expense of the energetically unfavorable AA regions.
In HTG, the analogous effect can now occur at the supermoir\'e scale.
Indeed, as we will now demonstrate, while the moir\'e scale lattice reconstruction is minor at $\theta\approx1.5^\circ$, the supermoir\'e scale $a_{mm}\sim 300-400$nm is well in the regime of severe supermoir\'e lattice reconstruction.

We model in-plane lattice relaxation in HTG using the configuration space method developed in Refs~\cite{zhu2020modeling,carr2018relaxation}:
the total intra- and inter-layer energy, with parameters extracted from ab initio theory, is minimized in configuration space which avoids issues associated with real space incommensurability. %
From this, we extract a real space map of the local shift field $\mathbf{u}_l(\mathbf{r})$ which indicates the in-plane displacement of the relaxed structure relative to the unrelaxed structure, for each layer $l$.  

In Fig. 1a, we show the AA stacking regions of adjacent layer pairs, labeled by AA$_{12}$ and AA$_{23}$, for the relaxed structure.
The dramatic effect of supermoir\'e lattice reconstruction is clearly visible by eye: large domains separated by a triangular network of domain walls.
Within each domain, the AA$_{12}$ and AA$_{23}$ regions come together to form the two sublattices of a periodic moir\'e-scale honeycomb lattice, as shown in Fig1b.  
These are characterized by a finite lateral shift $\v d = \pm \v \delta$ (defined later) between the two moir\'e sublattices;
since $\v d$ is opposite in two adjacent domains, a domain wall must form between them.  
Thus, while HTG nominally forms a supermoir\'e structure, it is energetically favorable to relax to large domains of locally periodic regions.
We use the term periodic moir\'e superlattice to refer to these periodic structures, and specifically those realized in the upwards and downwards pointing triangular domains as h-HTG and $\overline{\mathrm{h}}$-HTG, respectively.

On the atomic scale, the high symmetry stacking regions in h-HTG correspond to AAB, ABA, and BAA stacking regions (Fig1b), while in $\overline{\mathrm{h}}$-HTG they are ABB, BAB, and BBA.
The relaxed structure therefore completely avoids the energetically costly AAA stacking region.
It is interesting to contrast our results to that of alternating-twist trilayer graphene, $\theta_l=(0,\theta,0)$, where the A-twist-A configuration (which does contain an AAA region) was shown, using the exact same method and energetic parameters, to be favorable~\cite{carrUltraheavyUltrarelativisticDirac2020,turkel2022orderly}. 
Our results demonstrate that the favorable stacking configuration is not obvious \emph{a priori}, and depends on subtle energetic properties. 

The periodic structure of h-HTG can be understood from the fact that unrelaxed HTG is already very close to a periodic moir\'e superlattice.
In Fig. 1c, we illustrate the monolayer Brillouin zone (BZ) of each graphene layer.
The moir\'e BZ (mBZ) for each layer pair, the edges of which are determined by $\v  K_{2}-\v K_{1}$ and $\v K_{3}-\v K_{2}$, are incommensurate with each other as they are rotated by a small angle $\pm\theta/2$.
However, this incommensurability can be remedied by a minuscule uniform compression of the outer graphene layers (and/or dilation of the middle layer) by a factor $\lambda=\cos(\theta) \approx 0.9997$  for $\theta=1.5^\circ$.
The result is that the new $\v K_l$ points all lie along a vertical line, satisfying $\v K_{2}-\v K_1=\v K_3-\v K_2$, and therefore resulting in a periodic moir\'e superlattice.
We define the commensurate mBZ as shown in Fig1c, in which the $\v K_1$, $\v K_2$, and $\v K_3$ points fold to the $\kappa$, $\gamma$, and $\kappa^\prime$ points, respectively.

To verify that this is the correct picture, we obtain the local twist angle $\theta_l(\v r)=\theta_l+\sin^{-1}[\frac{1}{2}\nabla\cross \v u_l (\v r)]$ and uniform scaling factor $\lambda_l(\v r)=1+\frac{1}{2}\nabla\cdot \v u_l(\v r)$ of the relaxed HTG structure.
We then define the ``local moir\'e aperiodicity'' via $A(\v r) \equiv \sum_{l=1,3}|K_{lx}(\v r)/K_{2x}(\v r)-1|$, where $K_{lx}(\v r) = K\cos[\theta_l(\v r)]/\lambda_l(\v r)$ is the ``local $K_x$'', and $K=\frac{4\pi}{3 a_0}$.
$A(\v r)$ is zero if all three $K$ points lie on a line and non-zero otherwise.
The local moir\'e aperiodicity is plotted in the background of Fig1c, which shows that the large triangular domains have indeed relaxed to the locally periodic structure with $A(\v r)\approx 0$.
Thus, the physics within each domain is indeed described by the periodic moir\'e structure with the mBZ illustrated in Fig1a.

In the domain wall region and their intersection, $A(\v r)>0$ and is much larger than in the unrelaxed structure,
$A_0=2|1-\cos\theta|\approx0.68\times 10^{-3}$.
This implies that these regions (which contain the previously studied $\v d=0$ model~\cite{mora2019flatbands,popov2023magic}) actually relax \emph{away} from the periodic structure, and therefore appears more locally quasicrystalline~\cite{uri2023superconductivity}.

We remark that, although the moir\'e period $a_m$ and the domain size (determined by the unrelaxed supermoir\'e period) $a_{mm}$ considered thus far are both determined by $\theta$, they can in principle be tuned independently.
This is important as it means that the domain size can be controlled while keeping the local physics within each domain fixed.
By applying a small global uniform compression to the outer layers via $\lambda<1$, which may be possible via heterostrain engineering, the domain size 
$a_{mm}= a_0\lambda/(2|\lambda-\cos\theta|)$ quickly increases and diverges at $\lambda=\cos\theta$ at which point the entire system is a single domain. %
This single domain structure has lower elastic energy density due to the absence of domain walls, so we speculate that some degree of this may already occur naturally in finite systems.
Finally, we remark that our conclusion about the relaxed structure are qualitatively insensitive to details such as the precise ratio of intra- and inter-layer elastic energies, which is a potential tuning knob in comparing with experiment~\cite{turkel2022orderly}.

\section{Electronic structure}

Having established the importance of lattice relaxation on the resulting supermoir\'e structure, we now turn to the electronic structure within a h-HTG domain.
Rather than deriving a quantitative electronic model based on the relaxed structure~\cite{carr2019exact,guinea2019continuum}, which would contain many detail-dependent terms, we instead take an effective approach that captures the essential physics.
The starting point for our analysis is the Bistritzer-MacDonald continuum model generalized to three layers. For more than two layers we must take into account the displacements of the two moir\'e superlattices, $\v d_{t,b}$.
\begin{equation}
H_K = 
\begin{pmatrix}
-iv\v{\sigma}_{\theta}\cdot \v{\nabla} & T(\v r -\v d_t) & 0\\
T^\dag(\v r -\v d_t)& -iv\v{\sigma} \cdot \v{\nabla} &  T(\v r-\v d_b) \\
0 & T^\dag(\v r-\v d_b) & -iv\v{\sigma}_{-\theta}\cdot \v{\nabla}\\
\end{pmatrix}\label{eq:HK}
\end{equation}
where $\v{\sigma}_\theta = e^{-i\theta \sigma_z}(\sigma_x,\sigma_y)$ and
\begin{equation}
T(\v r) = w \begin{pmatrix} \kappa U_0(\v r) & U_{-1}(\v r) \\ U_{1}(\v r) & \kappa U_0(\v r) \end{pmatrix} 
\end{equation}
is the moir\'{e} tunneling between layers, with $U_l(\v r) = \sum_{n=0}^2 e^{\frac{2\pi i}{3} l n} e^{-i \v q_n \cdot \v r}$. The tunneling wavevectors are such that $q_{n,x} + i q_{n,y} = -ik_\theta e^{\frac{2\pi i}{3} n}$, where $k_\theta = 2K\sin\frac{\theta}{2}$. We will use $v = 1.03\times10^{6} \text{m/s}$ and $w = 105$meV, which we believe well models trilayer graphene at these twist angles capturing some degree of interaction-induced velocity renormalization~\cite{turkel2022orderly,uri2023superconductivity}.

The intra-sublattice tunneling strength is suppressed due to lattice relaxation and renormalization by $\kappa < 1$; while hard to estimate precisely\cite{parkerFieldtunedZerofieldFractional2021}, TBG studies \cite{ledwithTBNotTB2021,carr2018relaxation,carr2019exact,carrMinimalModelLowenergy2019,nam2017lattice,koshinoEffectiveContinuumModel2020,koshinoMaximallyLocalizedWannier2018,guinea2019continuum,dasSymmetrybrokenChernInsulators2021,vafekRenormalizationGroupStudy2020} suggest $\kappa\approx0.5-0.8$, and we therefore take a conservative estimate $\kappa=0.7$ for now.
The Hamiltonian for the $K^\prime$ valley can be obtained by time reversal symmetry, and spin degeneracy is implied.
This model has a moir\'{e} translation symmetry with reciprocal lattice vectors $\v b_{1,2} = \v q_{1,2} - \v q_0$ and lattice vectors $\v a_{1,2} = \frac{4\pi}{3k_\theta}(\pm \frac{\sqrt{3}}{2}, \frac{1}{2})$. The Bloch periodicity of $l$'th layer is given by $\psi_{\v k,l}(\v r + \v a) = e^{i (\v k - \v K_l) \cdot \v a} \psi_{\v k, l}(\v r)$, where $\v K_{1,3} = \mp  \v q_0 + \v K_2$ are the $\kappa$ and $\kappa'$ points of the mBZ and $\v K_2$ is the $\gamma$ point. 

Since we may always translate the entire system at the moir\'{e} scale, only $\v d = \v d_t - \v d_b$,
the offset between the moir\'{e} patterns, affects the spectrum of the continuum Hamiltonian $H_K$. 
While a generic $\v d$ breaks most crystalline symmetries, there is an approximate particle-hole-inversion symmetry $\mathcal{IC}$ which exchanges the top and bottom layers, multiplies the middle layer by $-1$, takes $\v r \to -\v r$, and anticommutes with the Hamiltonian. This symmetry is exact if we take $\sigma_{\pm \theta} \to \sigma$, which is a very good approximation for the small $\theta$ of interest here, and is easiest to see if one chooses $\v d_t = - \v d_b$.

\begin{figure}[t]
\includegraphics[width=0.5\textwidth]{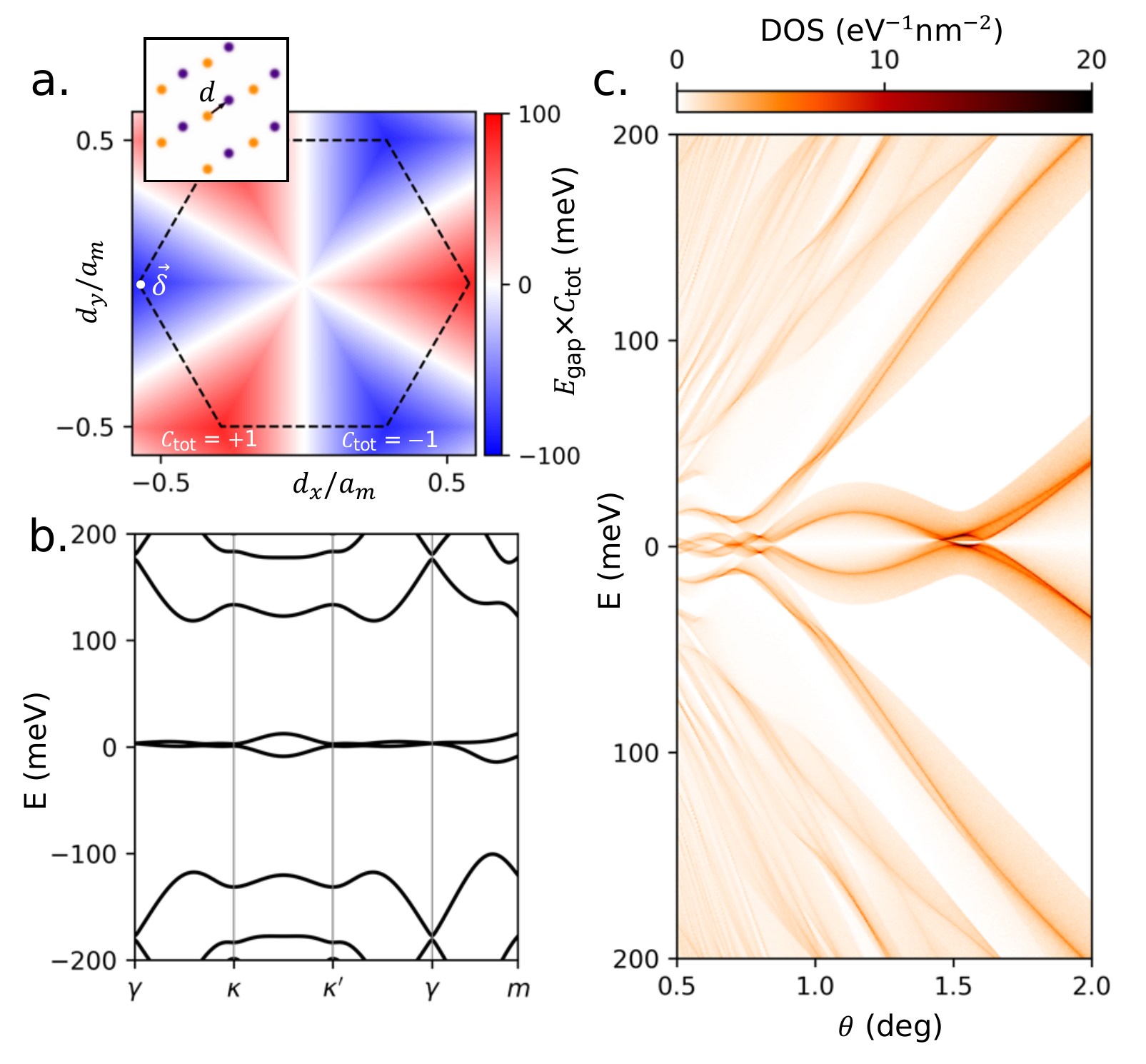}
\caption{(a) The remote band gap $E_{\mathrm{gap}}$ multiplied by $C_{\mathrm{tot}}=\pm1$ of $H_K$ is shown as a function of $\v d$, the relative offset between the moir\'e lattices (illustrated in inset), for $\theta=1.5^\circ$ and $\kappa=0.7$.
(b) The moir\'e band structure for h-HTG, corresponding to $\v d = \v \delta$.
(c) The density of states for h-HTG as a function of $\theta$.}
\end{figure}

In Fig2a, we show the remote band gap, defined as the minimum of the gap between the second and first conduction or valence bands, as a function of $\v d$ for $\theta=1.5^\circ$.
For special shifts such as $\v d=0$, or along high symmetry lines, the remote band gap is forced to be zero~\cite{mora2019flatbands,mao2023supermoire}.
For generic shifts, however, the remote band gap is non-zero and maximized for shifts at the corners of the moir\'{e} unit cell: $\v d=\pm \v \delta = \pm\frac{1}{3}(\v a_2 - \v a_1)$.
Computing the total Chern number of the first conduction and valence bands, we find $C_{\mathrm{tot}}=\mp 1$ in the regions smoothly connected to the high symmetry $\pm\v\delta$ points.
The corresponding bands in the $K^\prime$ valley have opposite Chern numbers by time-reversal symmetry.
The band structure at $\v d = \v \delta$, shown in Fig 2b, demonstrates the existence of two isolated nearly-flat bands carrying net topology.  %

We now focus on the properties of the h-HTG periodic moir\'e superlattice, obtained by setting $\v d = \v \delta$. 
The $\overline{\mathrm{h}}$-HTG model, obtained by setting $\v d = -\v \delta$, is related by $C_{2z}\mathcal{T} = \sigma_x \mathcal{K}$, where $\mathcal{K}$ is complex conjugation, which takes $\v r \to -\v r$ and leaves $\v k$ invariant. At this special value of $\v d$ the model has additional symmetries. Because $\v \delta$, as the corner of the unit cell, is a $C_{3z}$ invariant point, the resulting moir\'{e} superlattice is $C_{3z}$ symmetric. Furthermore, because $\v \delta \to -\v \delta$ under $x \to -x$, the model is additionally symmetric under $C_{2y}$, which exchanges the top and bottom layers and also exchanges valleys. The antiunitary mirror symmetry $C_{2y} \mathcal{T}$ acts within a valley. 

Fig2c shows the density of states (DOS) of h-HTG as a function of $\theta$.
The most prominent feature is the appearance of the advertised magic angle at $\theta\approx1.5^\circ$, where two topological flat bands appear at the charge neutrality point.
At the magic angle, the DOS exhibits a sharp peak ($\mathrm{DOS}>20$eV$^{-1}$nm$^{-2}$), the remote band gap is large $E_{\mathrm{gap}}\approx 85$meV, and the dispersion, half the total bandwidth of both bands, is small $W\approx 15$meV.
These values should be contrasted with a typical interaction scale; this can be obtained by scaling up the $20-30$meV estimate\cite{cao2018correlated,xieSpectroscopicSignaturesManybody2019} of TBG interactions by $\approx 1.5$, due to the larger angle, yielding the range $30-45$meV.
This ordering of energy scales is ideal for exploring strongly correlated topology, in which the large $E_{\mathrm{gap}}$ essentially ``locks in'' the quantum geometry of the flat band manifold, within which interactions are dominant.

\section{Chiral limit}

The origin of these flat bands can be understood from the chiral model\cite{tarnopolskyOriginMagicAngles2019}, obtained by setting $\kappa=0$, which we now analyze in detail.
Chiral models, with exactly flat bands\cite{tarnopolskyOriginMagicAngles2019,wangChiralApproximationTwisted2021,naumisReductionTwistedBilayer2021a,navarro-labastidaHiddenConnectionTwisted2022,renWKBEstimateBilayer2021,beckerSpectralCharacterizationMagic2021,beckerFineStructureFlat2022,beckerIntegrabilityChiralModel2022,shefferChiralMagicangleTwisted2021a,ShefferQueirozStern22}, also motivate a ``sublattice-Chern basis''\cite{bultinckGroundStateHidden2020,khalafChargedSkyrmionsTopological2021,ledwithStrongCouplingTheory2021}, useful for all $\kappa$, where the flat bands have Chern number $C=(1, -2)$. 
The chiral model enables strong coupling approaches\cite{ledwithStrongCouplingTheory2021} to several types of correlated insulating states including generalized ferromagnets, topological charge density waves, and fractional Chern insulators.

To write an explicit form of the chiral Hamiltonian, we choose $\v d_t = - \v d_b = -\v \delta$ such that $\v d = -2 \v \delta \equiv \v \delta$ and $U_l(\v r - \v d_{t,b}) = U_{l \pm 1}(\v r) $. We therefore have, in the basis where $\sigma_z = \text{diag}(1,1,1,-1,-1,-1)$,
\begin{equation}
\begin{aligned}
    H_K & = vk_\theta \begin{pmatrix} 0 & D^\dag \\ D & 0 \end{pmatrix}, \\
    D & = \begin{pmatrix} 
    -2i e^{i \zeta} \ov{\partial} & \alpha U_{-1}(\v r) &  0  \\
    \alpha U_0(-\v r) & -2i \ov{\partial} & \alpha U_0(\v r) \\
    0 & \alpha U_{-1}(-\v r) & -2i e^{-i \zeta} \ov{\partial}
    \end{pmatrix}.
    \end{aligned}
    \label{eq:chiralHam}
\end{equation}
Here we have nondimensionalized the Hamiltonian using $\v r \to \v r k_\theta$, $\ov{\partial} \to \ov{\partial}/k_\theta$ where $\ov{\partial} = \frac{1}{2}(\partial_x + i \partial_y)$. Nominally $\zeta = \theta$, but it can be instructive to imagine tuning it independently;
none of our conclusions depend on its precise value.

As we tune the dimensionless tunneling strength $\alpha = w/vk_\theta \sim 1/\theta$, we find a sequence of magic angles, listed in Table 1, at which we obtain \emph{exactly} flat bands at zero energy, seen by the vanishing bandwidth in Fig3a.
Due to the chiral symmetry of the Hamiltonian $\{H,\sigma_z\}=0$, we may label zero modes by their eigenvalue under $\sigma_z$, such that the flat bands correspond to zero modes of $D$ polarized on the A sublattice and of $D^\dag$ on the B sublattice. 
As shown in Fig3b, odd parity magic angles have two flat bands per spin per valley, while even-parity magic angles have four together with a dispersive Dirac cone at $\Gamma$ (for $\zeta=0$). %
Interestingly, the even magic angle dispersive cone is gapped out by $\zeta\neq 0$ but the four exactly flat bands remain. The distinction between even and odd magic angles, together with ratios between magic $\alpha$ that do not match those of TBG, suggest that the magic angles here do not descend from those of TBG. This is in contrast to the chiral magic angles of twisted chirally stacked multilayers\cite{ledwithFamilyIdealChern2022,khalafMagicAngleHierarchy2019}, alternating twist multilayers\cite{khalafMagicAngleHierarchy2019}, and the $\v d = 0$ periodic HTG~\cite{popov2023magic},  which can all be related to TBG.
A detailed understanding of the mathematical structure of this model is an interesting subject beyond the scope of this work.

Let us focus on the first magic angle $\alpha_1 \approx 0.377 + O(\zeta)$ which is the most experimentally relevant. 
Here we obtain two exactly flat bands; the A sublattice band has $C_A=1$ and the B sublattice band has $C_B=-2$.

To understand the emergence of flat bands and their Chern numbers we begin with the 3 Dirac cones associated with the $\alpha = 0$ decoupled limit. These cones are protected and pinned to zero energy by chiral symmetry, pinned to the $\kappa,\gamma,\kappa^\prime$ points by $C_3$ symmetry, and all have positive chirality. The net chirality of three implies that the flat bands obtained by gapping all three cones with a $\sigma_z$ mass results in bands that differ in Chern number by $C_A - C_B = 3$ \cite{gao2022untwisting}. The low-energy bands must therefore have a net topology $C_A + C_B \neq 0$. 

\begin{figure}
\includegraphics[width=0.5\textwidth]{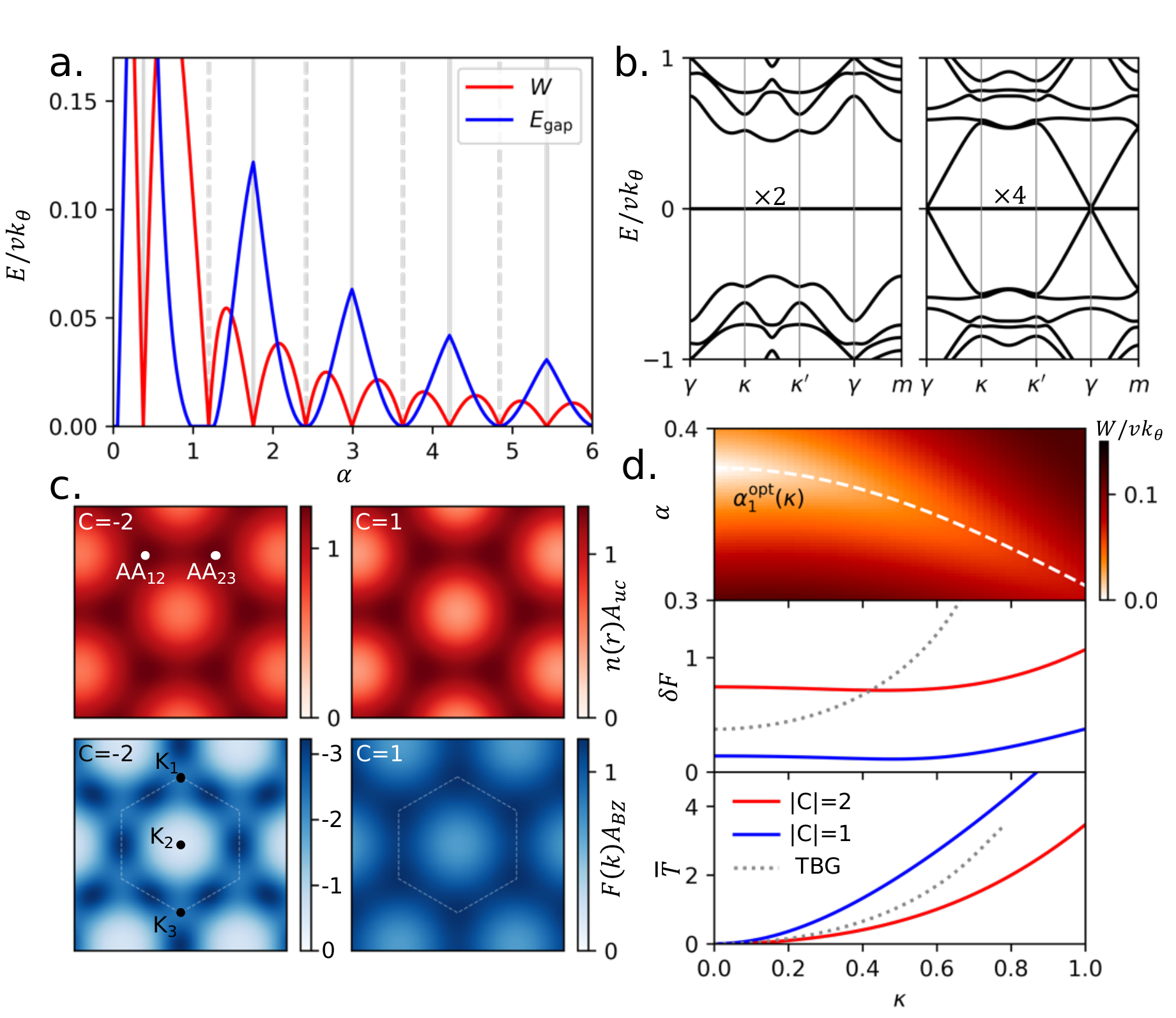}
\caption{
(a) The dispersion $W$ and remote gap $E_{\mathrm{gap}}$ for the chiral model with $\zeta=0$, as a function of the dimensionless tunneling parameter $\alpha$.
(b) The band structure at $\alpha_1$ (left) and $\alpha_2$ (right).
(c) The charge density $n(\v r)$ and Berry curvature $F(\v k)$ in the Chern basis for the $|C|=1,2$ bands, where $A_{uc}$ and $A_{BZ}$ are the moir\'e unit cell and mBZ areas respectively.
(d) The evolution of $W$, Berry curvature deviation $\delta F$, and trace condition violation $\overline{T}$, as a function of $\kappa$ in the Chern basis, and corresponding TBG values for comparison.
}
\end{figure}
\begin{table}
\begin{tabular}{|c|c|c|c|c|c|c|c|c|}
\hline
$\alpha_1$&
$\alpha_2$&
$\alpha_3$&
$\alpha_4$&
$\alpha_5$&
$\alpha_6$&
$\alpha_7$&
$\alpha_8$&
$\alpha_9$
\\
\hline
0.377&
1.297&
1.755&
 2.414&
 2.991&
 3.628&
 4.213&
 4.840&
 5.430\\
\hline
\end{tabular}
\caption{List of magic angles for the chiral model with $\zeta=0$.}
\end{table}

We now analytically derive the exact flatness and Chern numbers at the first magic angle.
Let us focus on the $A$ sublattice, $\psi = \psi_A$. From the decoupled limit, we find that the $C_{3z}$ representation of the zero mode $\gamma$-point wavefunction is such that $\psi_{\gamma 1,3}(\v r = 0) = 0$ but $\psi_{\gamma 2}(0)$ is in general nonzero. In a $C_{2y} \mathcal{T}$ symmetric gauge, $\psi_{\gamma 2}(0)$ is a signed real number, and it is natural for it to cross through zero \cite{ledwithStrongCouplingTheory2021,ShefferQueirozStern22}; we find that it does at $\alpha=\alpha_1$. 
Such a crossing point is stable to $C_{3z}$ and $C_{2y}\mathcal{T}$ preserving perturbations, and leads to exactly flat bands.
At the crossing point, the entire $\gamma$ point zero-mode wavefunction vanishes at $\v r=0$. We may therefore write\cite{tarnopolskyOriginMagicAngles2019,wangChiralApproximationTwisted2021}
\begin{equation}
    \psi_{\v k}(\v r) = e^{\frac{i}{2} \ov{k} z} \frac{\sigma(z+ik)}{\sigma(z)} \psi_\gamma(\v r)
    \label{flatAsublattice}
\end{equation}
as a zero mode wavefunction at wavevector $\v k$, measured from the $\gamma$ point for concreteness. Here, $\sigma(z) = \sigma(z|a_1,a_2)$ is the (modified \cite{haldaneModularinvariantModifiedWeierstrass2018}) Weierstrass sigma function which satisfies $\sigma(-z) = -\sigma(z)$ and $\sigma(z + a_{1,2}) = -e^{\frac{1}{2} \ov{a}_{1,2}\left(z + \frac{a_{1,2}}{2}\right)}$. The pole associated with the zero of the sigma function is cancelled by the zero of $\psi_\gamma$.
Here we have used the complex number notation $k = k_x + i k_y$, $z=x+iy$, and $a = a_x + i a_y$. We note that \eqref{flatAsublattice} may be interpreted as the wavefunction of a Dirac particle moving in an effective inhomogeneous magnetic field of $2\pi$ flux per unit cell, where $B_{\rm eff} (\v r) = \nabla^2 \log \frac{\abs{\sigma(z)}}{||\psi_\gamma(\v r)||} $\cite{ledwithFractionalChernInsulator2020,wangExactLandauLevel2021}, up to an unimportant $\v k$-independent normalized layer vector.

The fact that $\psi$ has a single $k$-space zero, for each $\v r$, of positive winding implies that the band has $C_A=1$ (there is a winding by $2\pi$ around the mBZ)\cite{wangExactLandauLevel2021}. It is also possible to compute the Chern number from the $k$-space quasiperiodicity of \eqref{flatAsublattice} \cite{ledwithStrongCouplingTheory2021,wangExactLandauLevel2021}. Since $D$ and $D^\dag$ have the same singular values when there is no external magnetic flux\cite{shefferChiralMagicangleTwisted2021a}, the B sublattice must also have an exact band of zero modes from which $C_A - C_B = 3$ implies $C_B=-2$.

In Fig3c, we show the charge density $n(\v r)$ and Berry curvature $F(\v k)$ for these bands.
The charge distribution for both bands are remarkably uniform.
This can be motivated from the fact that TBG has a singly peaked charge density at the AA region:
since h-HTG contains two such AA sites, AA$_{12}$ and AA$_{23}$, the charge density has two peaks at these locations for each layer pair and is overall much more uniform. 
We also find that the Berry curvature distribution for both bands feature a multi-peak structure in momentum space, and are relatively uniform (with the A sublattice $C=1$ band being extremely uniform).
These are both important features reminiscent of the lowest Landau level which persist to larger $\kappa$, as we discuss later.

All the Chern basis bands, labeled by valley, spin, and sublattice, are explicitly summarized in Table 2.

\begin{table}
\begin{tabular}{|c|l|l|}
\hline
Band & h-HTG & $\overline{\mathrm{h}}$-HTG \\
\hline
$(K,s,A)$ & $C=1$ & $C=2$\\
$(K,s,B)$ & $C=-2$ & $C=-1$\\
$(K^\prime,s,A)$ & $C=-1$ & $C=-2$\\
$(K^\prime,s,B)$ & $C=2$ & $C=1$\\
\hline
\end{tabular}
\caption{Table of Chern basis bands labeled by (valley, spin, sublattice), showing their Chern numbers, in the h-HTG and $C_{2z}$-related $\overline{\mathrm{h}}$-HTG structure.}
\end{table}

\section{Correlated states}

We now discuss the correlated physics and band geometry of these bands. 
At integer filling, generalized quantum Hall ferromagnets, obtained by filling any combination of bands within the Chern basis (Table 2), are very energetically competitive; they are exact eigenstates in the chiral limit, and exact zero energy ground states when bare and Hartree dispersions can be neglected\cite{repellinFerromagnetismNarrowBands2020,bultinckGroundStateHidden2020}. 
At filling $\nu = \pm 3$, measured relative to charge neutrality such that the empty and full flat bands have $\nu = \mp 4$ respectively, the generalized ferromagnetic state necessarily carries non-zero Chern number, though we will later discuss other states which avoid this restriction.
Coherence between bands of differing Chern number is heavily penalized because such order parameters must have vortices in the Brillouin Zone, equal in number to the difference in Chern number\cite{bultinckMechanismAnomalousHall2020}. 
Since no two Chern numbers are the same (including the other valley we have $C=\pm1, \mp 2$), we therefore do not expect intervalley coherence (IVC) in this system. Indeed, while TBG has an approximate $\mathrm{U}(4) \times \mathrm{U}(4)$ symmetry\cite{bultinckGroundStateHidden2020,VafekKangSymmetry,bernevigTBGIIIInteracting2020} that can rotate between IVC and valley diagonal orders, we have the $\mathrm{U}(2) \times \mathrm{U}(2) \times \mathrm{U}(2) \times \mathrm{U}(2)$ subgroup consisting of spin and charge rotations for each sublattice and each valley. For nonzero bare dispersion, or outside the chiral limit, this symmetry is broken to $\mathrm{U}(2) \times \mathrm{U}(2)$ consisting of spin and charge rotations in each valley.
A rich phase diagram of spontaneous symmetry breaking has been observed in TBG~\cite{zondinerCascadePhaseTransitions2020,wongCascadeElectronicTransitions2020,yu2022correlated}, and we expect similar physics to arise here.  
In this sense, the Chern basis is meaningful and important even outside the chiral limit \cite{bultinckGroundStateHidden2020,chatterjeeSymmetryBreakingSkyrmionic2020,ledwithStrongCouplingTheory2021}; to access it one can diagonalize the band projected sublattice operator $\Gamma_{\alpha \beta} = \bra{u_{\v k \beta}}\sigma_z \ket{u_{\v k \alpha}}$, where $\ket{u_{\v k \alpha}}$ is the Bloch wavefunction at wavevector $\v k$ associated to band $\alpha$. 

Zero mode bands of chiral Hamiltonians of the form \eqref{eq:chiralHam} have ``ideal quantum geometry"\cite{ledwith2022vortexability,ledwithFamilyIdealChern2022,ledwithFractionalChernInsulator2020,dong2022exact,wangExactLandauLevel2021,ledwithStrongCouplingTheory2021,parkerFieldtunedZerofieldFractional2021,gao2022untwisting,varjas2022} for fractional Chern insulators in a sense that we now describe. 
Because $D$ only depends on antiholomorphic derivatives, the zero mode band of $D$ maps to itself under multiplication by $z = x+iy$; we have $zP = PzP$
where $P$ is the projector onto the band and $z = x+iy$ can be thought of as a vortex operator; this condition is referred to as ``vortexability''\cite{ledwith2022vortexability}; vortices may be added while remaining within the band of interest. This condition may be iterated to replace $z$ with any holomorphic function $f(z)$. In momentum space, vortexability is equivalent to the ability to choose a gauge where the wavefunctions $u_k = e^{-i \v k \cdot \v r} \psi_{\v k}$ are holomorphic in $k_x + i k_y$ \cite{meraKahlerGeometryChern2021,ozawaRelationsTopologyQuantum2021,meraEngineeringGeometricallyFlat2021,leeBandStructureEngineering2017,claassenPositionmomentumDualityFractional2015,ledwith2022vortexability}. It is also equivalent to the momentum space ``trace condition"\cite{royBandGeometryFractional2014,liuRecentDevelopmentsFractional2022,parameswaranFractionalQuantumHall2013,jacksonGeometricStabilityTopological2015}, the saturation of the inequality $\overline{T} = \int d^2 \v k (\tr g_{\text{FS}}(\v k) - |F(\v k)|) \geq 0 $, where $g_{\text{FS}}$ is the Fubini-Study metric and $F(\v k)$ is the Berry curvature.
We say a system has ideal quantum geometry if $\overline{T} = 0$.

Ideal quantum geometry is intimately related to fractional Chern insulator ground states; we may begin with an ordinary many-body state $\ket{\Psi_0}$, e.g. the fully filled state, and create the state \cite{ledwithFamilyIdealChern2022,ledwith2022vortexability,dong2022exact}
\begin{equation}
    \ket{\Psi_{2s}} = \prod_{i<j} (z_i - z_j)^{2s} \ket{\Psi_0}
    \label{eq:manybody_vortexattach}
\end{equation}
which lies entirely within the band of interest due to the vortexability condition. This construction generalizes that of the $\nu = 1/(2s+1)$ Laughlin state but also applies to bands with $C>1$. 
If the band is flat and the band-projected interaction is sufficiently short-ranged and normal ordered with respect to an empty ``vacuum'', then $\ket{\Psi_{2s}}$ is the unique ground state at its filling factor\cite{haldaneFractionalQuantizationHall1983,trugmanExactResultsFractional1985,ledwithFractionalChernInsulator2020,ledwith2022vortexability,wangExactLandauLevel2021}. 

The previously mentioned charge density and Berry curvature homogeneity further help with stability to long-ranged interactions, which can be motivated by analogy with the lowest Landau level~\cite{royBandGeometryFractional2014,jacksonGeometricStabilityTopological2015,devakul2021magic}.
Additionally, the interaction generated Hartree dispersion obtained by integrating out already-filled bands is much smaller if charge density is peaked at more than one point\cite{gao2022untwisting}.
This is indeed the case here, especially relative to TBG where an AA-peaked charge density leads to a strong Hartree dispersion that works against FCIs \cite{parkerFieldtunedZerofieldFractional2021}.
Moving away from the chiral limit, we show the evolution of various geometric indicators in Fig3d.
We first observe the dispersion $W$ increasing with $\kappa$, and we identify the optimal magic angle $\alpha_1^{\mathrm{opt}}(\kappa)$ by the minimum in $W$.
We show the Berry curvature deviation $\delta F=\left(\int d\v k [\frac{1}{2\pi}F(\v k)-C]^2\right)^{\frac{1}{2}}$ and trace condition violation $\overline{T}$ of the two bands in the Chern basis, at $\alpha_1^{\mathrm{opt}}(\kappa)$, as a function of $\kappa$, along with the corresponding values calculated for TBG.
We find that $\delta F$ shows remarkably weak dependence on $\kappa$, and is significantly lower than TBG for both bands at realistic values of $\kappa$.
Turning to $\overline{T}$, we find that the $|C|=2$ ($1$) band is uniformly more (less) ideal than TBG, but are all of similar magnitude.
Overall, these favorable quantum geometric indicators are highly suggestive of an FCI ground state at fractional filling, and call for a detailed numerical analysis. 

Many correlated states have been predicted for $C=2$ bands, from the starting point of ideal geometry~\cite{dong2022exact,wang2022origin}. By doubling the unit cell, the $C=2$ band may be split~\cite{wuBlochModelWave2013,Kumar2014Generalizing,barkeshliTopologicalNematicStates2012} into two $C=1$ bands that are each individually vortexable~\cite{dong2022exact,wang2022origin} and related by translation symmetry; linear combinations of these $C=1$ subbands lead to a nearly-degenerate manifold of charge and spin density waves that can occur at half-integer fillings~\cite{dong2022exact,wang2022origin}. These states are guaranteed to be stabilized in the limit of short-ranged interactions when the bands have ideal geometry, similar to FCIs, but are numerically present for realistic parameters as well~\cite{wilhelmInterplayFractionalChern2021}. A $C=1$ insulator of this nature at half-integer filling was observed in twisted monolayer-bilayer graphene~\cite{polshyn2020electrical}.
We expect that this could be the case in h-HTG as well. By filling two of the $|C|=1$ sub-bands, say in opposite valleys, it is possible to also obtain integer filling states with unexpected properties, such as a $\nu = 3$ insulator with net Chern number zero. 
At fractional filling, there are a variety of fractional Chern insulating states that have been proposed in higher Chern bands~\cite{wuBlochModelWave2013,behrmannModelFractionalChern2016,wangFractionalQuantumHall2012,liuFractionalChernInsulators2012,trescherFlatBandsHigher2012,yangTopologicalFlatBand2012,sterdyniakSeriesAbelianNonAbelian2013,andrewsStabilityFractionalChern2018,andrewsStabilityPhaseTransitions2021,barkeshliTopologicalNematicStates2012,barkeshliTwistDefectsProjective2013}. Many can be constructed from \eqref{eq:manybody_vortexattach} using different parent states $\ket{\Psi_0}$. We refer readers to Refs. \cite{dong2022exact,wang2022origin} for more details.

We briefly discuss the incommensurate Kekul\'e spiral state, which was recently found to be important in strained TBG~\cite{kwan2021kekule,wagner2022global,wang2022kekul,nuckolls2023quantum}.  We expect that such incommensurate orders are less likely in h-HTG since their energetics appear to rely on a large, peaked, Hartree dispersion. The large Hartree dispersion originates from particular features of TBG wavefunctions, such as a single peak of the charge density per unit cell, that are not shared by h-HTG. Instead, we expect the previously discussed topological states to be favored.

We have highlighted that the topological nature and nearly ideal quantum geometry of h-HTG lead to a wide variety of potential correlated states, from generalized quantum Hall ferromagnets to topological density waves to fractional Chern insulators. 
Due to the proximity to the $\kappa = 0$ chiral limit, there is a pathway towards understanding the competition between these various correlated states. 
The investigation of the detailed energetic competition of such correlated states is an important subject for future works. 

\section{Discussion}

Having discussed extensively the rich correlated physics of h-HTG, we now briefly discuss the global physics when the domain size is finite.  
In this case, HTG separates into large domains of h-HTG and $\overline{\mathrm{h}}$-HTG, related by the valley-preserving $C_{2z}\mathcal{T}$ transformation which flips the sign of all Chern numbers.
When the domains are tuned to incompressible fillings, the low energy physics is dominated by the network of domain wall states which provide a realization of Chalker-Coddington type network models~\cite{chalker1988percolation,ho1996models}.
At the full or empty band insulating states $\nu=\pm 4$, stabilized by the $E_{\mathrm{gap}}\sim100$meV remote gap, each domain is a quantum valley Hall state with net valley-Chern number that is opposite in adjacent domains: we therefore expect the appearance of gapless edge modes counter-propagating along the network of domain walls.
Similar physics of counter-propagating edge networks has been actively explored in marginally twisted $\theta \ll 1^\circ$ TBG
~\cite{ yooAtomicElectronicReconstruction2019,huangTopologicallyProtectedHelical2018,rickhausTransportNetworkTopological2018,xuGiantOscillationsTriangular2019,liuTunableLatticeReconstruction2020,mahapatraQuantumHallInterferometry2022,renRealSpaceMappingLocal2023, chenCorrelatedStatesTriangular2020, chouHofstadterButterflyFloquet2020, debeuleAharonovBohmOscillationsMinimally2020, debeuleNetworkModelFourterminal2021, efimkinHelicalNetworkModel2018, houMetallicNetworkTopological2020,san-joseHelicalNetworksTwisted2013, tsimPerfectOnedimensionalChiral2020, wittigLocalizedStatesCoupled2023}, where a strong vertical displacement field is needed to gap the AB and BA domains.
In HTG, the large intrinsic band gap eliminates the need for a displacement field.
More interesting possibilities also arise here due to the correlated topological states.
When the system is valley polarized into QAH domains, adjacent domains carry opposite Chern numbers thus realizing a network of gapless chiral edge modes.
Alternatively, adjacent domains may have opposite valley polarizations; the competition between these possibilities depends on detailed domain wall energetics~\cite{kwan2021domain} and may be probed by an out-of-plane magnetic field.
This situation is similar to the Chern mosaic in hBN-aligned magic-angle TBG~\cite{shi2021moire,grover2022chern}.
At fractional filling, an even more exotic possibility is a network of FCI edge states.
A detailed theory of the emergent gapless networks and their interplay with the correlated topological states in HTG is an important subject for future works.

While we have mainly discussed HTG in the context of TBG, 
flat Chern bands have also been studied in a variety of other $C_{2z}$-breaking graphene structures such as twisted chirally-stacked multilayers\cite{zhangNearlyFlatChern2019,liuQuantumValleyHall2019,haddadiMoireFlatBands2020,zhangChiralDecompositionTwisted2020,maMoirBackslashFlat2020,chebroluFlatBandsTwisted2019,liu2020tunable,ledwithFamilyIdealChern2022,wangHierarchyIdealFlatbands2022,dong2022exact,wang2022origin}  and periodically strained graphene\cite{debeuleNetworkModelPeriodically2023,manescoCorrelationsElasticLandau2021, milovanovicBandFlatteningBuckled2020, phongBoundaryModesPeriodic2022,gao2022untwisting}.
Twisted monolayer-bilayer graphene in particular also has $C = \pm(1,-2)$. All of these systems have natural ``chiral limits"\cite{tarnopolskyOriginMagicAngles2019} where the band-wavefunctions have ideal quantum geometry for fractional Chern insulators \cite{ledwithFamilyIdealChern2022,wangHierarchyIdealFlatbands2022,dong2022exact,wang2022origin,gao2022untwisting}.
However, for monolayer-bilayer and bilayer-bilayer the exact chiral limit requires ignoring trigonal warping, and in its presence a large displacement field must be added to flatten the bands and reveal correlated states in experiment \cite{liu2020tunable,cao2020tunable,xu2021tunable,chen2021electrically,he2021competing,polshyn2020electrical,heChiralitydependentTopologicalStates2021,he2021symmetry}. In total, these realities likely impair the ideal quantum geometry for fractional Chern insulators in these systems. While strained graphene's chiral-flat $\abs{C}=1$ band is more robust~\cite{gao2022untwisting}, realizing the strain requires nanorod engineering~\cite{jiangVisualizingStrainInducedPseudomagnetic2017} or buckling over a $C_{2z}$-breaking substrate~\cite{maoEvidenceFlatBands2020}; the latter has only been achieved with the metallic $\text{NbSe}_{\text{2}}$~\cite{maoEvidenceFlatBands2020}, which precludes tuning the density with a gate. 
Furthermore, HTG is unique in its supermoir\'e scale domain walls and gapless edge-modes between $C_{2z} \mathcal{T}$ related topological states.

We have therefore demonstrated that HTG is a unique and exciting platform for realizing strongly correlated topological states, without the need for substrate alignment.
While unrelaxed HTG forms a moir\'e-of-moir\'e pattern, we have shown that lattice relaxation favors the formation of h-HTG, a periodic $C_{2z}$-breaking single-moir\'e superlattice.
We identify the relevant continuum model description for h-HTG and identify a magic angle at which a pair of flat topological bands with near-ideal quantum geometry isolated by a large remote band gap emerges.  
These flat bands can be traced back to a chiral limit, which enables a controlled starting point for a strong coupling approach to correlated topological states.
Our work lays the foundation for future theoretical and experimental studies of the strongly correlated topological physics in this platform.

\acknowledgements
We thank Ziyan Zhu for useful discussions and for collaboration on a related project. TD thanks Yves Kwan for helpful discussions.  PJL thanks Eslam Khalaf, Ashvin Vishwanath, Daniel Parker, Junkai Dong, Grigory Tarnopolsky, and Qiang Gao for collaborations on related projects. This
work was supported by the Air Force Office of Scientific Research (AFOSR) under award FA9550-22-1-0432. 
This work was partially supported by the Army Research Office MURI W911NF2120147, the 2DMAGIC MURI FA9550-19-1-0390, the National Science Foundation (DMR-1809802), the STC Center for Integrated Quantum Materials (NSF grant no. DMR-1231319), and the Gordon and Betty Moore Foundation’s EPiQS Initiative through grant GBMF9463 to PJH. AU acknowledges support from the MIT Pappalardo Fellowship and from the VATAT Outstanding Postdoctoral Fellowship in Quantum Science and Technology.

\bibliography{ref.bib}

\end{document}